# Observation of High-Order Harmonic Resonances in Magnetooptical Measurements of (BEDT-TTF)$_2$Br(DIA)

Yugo OSHIMA[1], Hitoshi OHTA[2], Keiichi KOYAMA[3], Mituhiro MOTOKAWA[3],

Hiroshi M. YAMAMOTO[4] and Reizo KATO[4]

[1]*The Graduate School of Science and Technology, Kobe University, Rokkodai 1-1, Nada, Kobe 657-8501*

[2]*Molecular Photoscience Research Center, Kobe University, Rokkodai 1-1, Nada, Kobe 657-8501*

[3]*Institute for Materials Research, Tohoku University, Katahira, Sendai 980-8577*

[4]*RIKEN (The Institute of Physical and Chemical Research), Hirosawa, Wako 351-0198*

Abstract

Magnetooptical measurements of a quasi-two-dimensional (q2D) organic conductor (BEDT-TTF)$_2$Br(DIA) (DIA=diiodoacetylene) were performed using a cavity perturbation technique with a millimeter vector network analyzer (MVNA). Harmonic resonances were observed periodically as a function of the inverse field up to the 7th order. This is the first observation of higher-order harmonics up to the 7th order in an organic conductor. The observed resonance is assigned as a periodic orbit resonance (POR) which arises from a q2D elliptic Fermi surface (FS). The obtained effective mass ($4.7m_e$) is consistent with Shubnikov-de Haas measurements. Temperature dependence of the spectra observed in two different sample configurations will be presented.

Quasi-two-dimensional (q2D) organic conductors (BEDT-TTF)$_2$X, where X stands for anions, have attracted considerable interest, because their physical properties change with various anions. For this reason, the Fermi surface (FS) topologies of these salts have been studied by various high magnetic field techniques, including measurements of de Haas-van Alphen (dHvA) oscillations, Shubnikov-de Haas (SdH) oscillations and angle-dependent magnetoresistance oscillations (AMRO). Another useful technique is magnetooptical measurements, and a few examples have been performed for the organic conductor.[1-9] For example, α-(BEDT-TTF)$_2$KHg(SCN)$_4$ is a very interesting organic conductor. It has an antiferromagnetic order below 12 K which is thought to be caused by a spin density wave (SDW) formation.[10] Although a recently proposed B-T phase diagram based on magnetic torque measurements or NMR measurements favors a charge density wave (CDW) scenario,[11,12] there still exists no clear evidence for a lattice modulation. Therefore, the ground state of this phase is quite complex and is still under discussion. We have reported magnetooptical measurements of this salt and observed four cyclotron resonances (CR) in which each effective mass was smaller than those obtained by dHvA and SdH oscillation measurements,[7] and this difference can be explained by considerable electron-electron interaction in the system (Kohn's theorem).[13,14] However, the FS of this salt is very complex at this temperature. Therefore, we could not clarify which resonance corresponds to the FS's closed orbit. Thus, our strategy is to attempt magnetooptical measurements of organic conductors which have a simple FS to determine the relationship between the cyclotron mass and Kohn's theorem. Moreover, recent magnetooptical measurements studies have shown that the mechanisms responsible for resonant absorption are quite different from conventional CR. There are two interesting predictions of harmonic resonances which are based on a simple FS structure.[15,16] One is a harmonic cyclotron resonance (HCR),[15] and the other is a periodic orbit resonance (POR).[16] HCR predicts the presence of high-order CRs which are associated with the higher harmonics of oscillating real-space velocity of charge carriers in cyclotron orbits around the FS pockets, which does not predict even harmonics. POR is a resonance coming from the conductivity along the least conducting direction, which arises from periodic motion in a plane perpendicular to the applied magnetic field where second and third harmonics are predicted. Therefore, conductors with a simple FS are advantageous to observe these predicted new types of resonances. (BEDT-TTF)$_2$Br(DIA) (DIA = diiodoacetylene) is a new q2D conductor with a very simple FS synthesized by Yamamoto et al.[17] One of the interesting features of the structure is that the supramolecular ...Br...DIA... chains are formed. The donor molecules fit into the channels formed by the one-dimensional chains along the a-c direction. SdH and AMRO measurements have already been performed by Uji et al. and the results indicate the presence of q2D FS with an elliptic cross-sectional area.[18] We will report our results of magnetooptical measurements of (BEDT-TTF)$_2$Br(DIA) which has a simple electronic structure and discuss the obtained effective mass in this paper.

We performed magnetooptical measurements using a cavity perturbation technique with a millimeter vector network analyzer (MVNA) at IMR, Tohoku University. The experimental setup can be found in ref. 19 and we refer the reader to a series of articles on the cavity perturbation technique.[20-24] The basic principle of the cavity perturbation technique can be described as follows. Provided that a sample placed inside a resonant cavity acts as a small perturbation of the electromagnetic field distribution within the cavity, we can determine the complex electrodynamic response of the sample from the changes of the quality factor of the resonance ($Q$) and the resonance frequency ($\omega_0$). For a conducting sample, changes in quality factor and resonance frequency are directly related to changes in the surface impedance $Z_s=R_s+iX_s$, where $R_s$ is a surface resistance and $X_s$ is a surface reactance. The dissipation, which occurs at the surface of the sample due to $R_s$, generally causes changes in quality factor $Q$, while the dispersion, governed by $X_s$, results in changes in resonance frequency $\omega_0$. The amplitude "$A$" and phase "$\phi$" of the electric and magnetic fields of the electromagnetic waves in the cavity can be written as follows.

$$|A(\omega)|^2 = \frac{A_0^2}{(\omega_0/2Q)^2 + (\omega-\omega_0)^2} \quad (1)$$

and

$$\phi = -\arctan\left(\frac{2(\omega - \omega_0)}{\omega_0/2Q}\right) \quad , \qquad (2)$$

where $A_0$ is the notional amplitude of the field supplying energy to the cavity and $\omega$ is the frequency used. Therefore, we used MVNA to monitor the phase and amplitude of millimeter-wave radiation transmitted through a resonant cavity containing the sample under investigation. The typical sample size used for this study was 1x1x0.1 mm$^3$ and the static magnetic field was always applied perpendicular to the conducting plane ($H//b^*$). Cylindrical cavities were used for this study and the frequencies used are around 58 GHz and 72 GHz whose cavity mode corresponds to TE$_{011}$ and TE$_{012}$, respectively. We performed two sample positions for this measurement (pillar and end-plate configurations). In the end-plate configuration, the sample was set on the end-plate (Fig. 1(a)) and notice that the mode coupling is always $H$ field coupling. In the pillar configuration, the sample was set in the middle of the cavity using a polyethylene pillar which had been set in the opposite side of the coupling hole (Fig. 1(a)). This configuration has a different coupling mode by using different frequency. We have an $E$ field and $H$ field couplings when using 58 GHz (TE$_{011}$) and 72 GHz (TE$_{012}$), respectively. For the $H$ field coupling, an induced current rounds the sample (Fig. 1(b)). For the $E$ field coupling, the currents will flow within the plane surface from the sample edges and faces (Fig. 1(c)) (See discussion in ref. 25). There are always skin-depth effects in high-frequency measurements for conducting materials. However, the in-plane and interlayer conductivities differ considerably (the interplane conductivity is 3 to 4 orders of magnitude smaller) for q2D BEDT-TTF salts. Thus, the interplane skin depth ($\delta_\perp$) will be larger than the skin depth for in-plane ac currents ($\delta_{//}$). This is why so many papers suggest that the interplane ac magnetoconductivity is mainly probed for this type of measurement.[16,23-25] The in-plane and interplane dc conductivities of (BEDT-TTF)$_2$Br(DIA) are around 10$^3$ and 1-10 S/cm, respectively, which are relatively lower than those of other BEDT-TTF salts. These correspond to 6.5 and 65-210 μm of in-plane ($\delta_{//}$) and interlayer ($\delta_\perp$) skin depths, respectively, when using 60 GHz. Regarding the relative contributions of the dissipation within the sample, the dissipation is governed by $R_s$ which is proportional to the skin depth. This means that the ratio of the power dissipation due to interlayer ($P_\perp$) and in-plane ($P_{//}$) can be written as $P_\perp/P_{//}=a_\perp\delta_\perp/a_{//}\delta_{//}$, where $a_\perp$ and $a_{//}$ are appropriate areas for the surface across which current flows.[25] Thus, the power dissipation ratio for this sample would be around 1-3.2, which suggests that interplane conductivity will be mainly probed, but we have also the possibility to observe both in-plane and interplane ac conductivities.

Figure 2(a) shows typical spectra of (BEDT-TTF)$_2$Br(DIA) at 0.5 K in the pillar configuration using several frequencies. This is the sample configuration when we successfully observed conventional CR with another q2D organic conductor, θ-(BEDT-TTF)$_2$I$_3$.[9] At each frequency, we can observe several harmonic absorptions which become larger as the field increases. We show in Fig. 2(b), the inverse-field plot of the spectrum at 72 GHz (TE$_{012}$). It is clear that the harmonic absorption lines appear periodically as a function of the inverse field. If the resonances are HCR or POR, the resonance condition will be $\omega=n\omega_c$, where $\omega_c(=eB/m^*)$ and $\omega(=2\pi\nu)$ are the cyclotron and microwave frequencies, respectively, which cause $n$th-order higher harmonics ($n$ will be an odd number if HCR is observed.).[15,16] We point out that these resonances are different from Azbel'-Kaner CR or Gor'kov-Lebed' CR,[26,27] because the skin depth of organic conductors is relatively larger than the cyclotron radius (e.g., ~1 μm for 60 GHz). From the condition mentioned above, the resonance periodicity against the inverse field should be described as follows.

$$\Delta\left(\frac{1}{B}\right) = \left(\frac{1}{B_{n+1}} - \frac{1}{B_n}\right) = \frac{e}{2\pi\nu m^*} \quad , \qquad (3)$$

where $B_n$ is the resonance field of $n$th-order harmonics and $m^*$ is the cyclotron effective mass. Therefore, from the periodicity of the resonances in Fig. 2(b), the estimated effective mass would be around 4.4$m_e$. Then, the order of higher harmonics would be given by

$$n = \frac{\omega}{\omega_c} = \frac{2\pi\nu m^*}{eB} \quad . \qquad (4)$$

Therefore, the strongest absorption corresponds to the second harmonic; we have observed harmonics up to the 7th order (n=1 resonance was not observed in this configuration). We note that this is the first observation of higher order harmonics in the field of organic conductors. We consider that the sample condition (i.e., skin depth, sample size) was appropriate for observing higher harmonics. Figure 3 shows the temperature dependence (from 0.7 K to 4.2 K) of the spectra around 58 GHz ($TE_{011}$) in the pillar configuration. The harmonics become more apparent as the temperature decreases. The effective mass estimated from periodicity is around $4.6m_e$. We do not think the broad absorption seen around 13 T at 0.7 K is intrinsic, because the peak disappeared with increasing temperature and n=1 resonance should appear at around 9.5 T. The harmonics up to the 6th order were observed at 58 GHz and the second harmonic still remains the strongest resonance. Therefore, we conclude that the observed resonances are POR, since the second harmonic is expected to be the strongest for POR.[16] It remains an open question why conventional CR has not been observed in this configuration. POR is a result of multiple cyclotron-resonance-like features in the conductivity along the least conducting direction ($\sigma_\perp$ in Figs. 1(b) and 1(c)).[16] Thus, the lineshape for POR should be sensitive to the electrodynamics around the sample. However, there is not a significant difference in the lineshapes at 58 GHz and 72 GHz in Fig. 2(a), while the induced currents are completely different with different frequencies (i.e., different coupling modes). This may be due to the ambiguity of the coupling mode when setting the sample in the middle of the cavity. Therefore, to clarify this problem, we performed measurements in the end-plate configuration whose coupling mode will always be *H* field coupling.

Figure 4 shows the typical spectra at 0.5 K in the end-plate configuration using different frequencies. Each spectrum was normalized so that the horizontal axis corresponds to the cyclotron mass. In general, the *H* field coupling mode is the ideal configuration for observing ESR signals. Thus, ESR signals are clearly observed in this configuration, which confirms the ambiguity of the coupling mode in the previous configuration. The strong absorption at $2.7m_e$ corresponds to the second harmonic; the n=1 resonance is observed at around $4.7m_e$, which has not been observed in the pillar configuration. The linewidth of the second harmonic in the end-plate configuration is larger than that in the pillar configuration. This may suggest that the interlayer conductivity ($\sigma_\perp$ in Fig. 1(b)) is dominant in the end-plate configuration, and that a mixture of interlayer and in-plane conductivity ($\sigma_\perp$, $\sigma_\parallel$) is observed in the pillar configuration. However, a higher order of harmonics (i.e. 5th, 6th, 7th) was not observed in this configuration. The in-plane conductivity $\sigma_\parallel$ might play an important role for the observation of the higher order of harmonics.

Figure 5 is the inverse-field plot of the spectrum shown in Fig. 4 (Note that the horizontal axis is normalized by *n*, using the cyclotron mass value estimated by the periodicity of the harmonics, $4.4m_e$ and $4.8m_e$ for 58 GHz and 72 GHz, respectively). Due to the difference between the direct cyclotron mass value and the estimated mass from periodicity, there is a slight deviation of the second harmonic in Fig. 5. We could not determine whether the n=1 resonance is a fundamental POR or a conventional CR. In the pillar configuration, no conventional CR was observed; thus, this may suggest that n=1 resonance is a fundamental POR. However, the problem still remains why it was not observed in the pillar configuration. If it is a fundamental POR, the observation of n=1 resonance may due to the sample misalignment or a tilting of the FS warping plane away from the conducting plane.[16] The reason why the POR is observed in $(BEDT-TTF)_2Br(DIA)$ while the conventional CR is observed in θ-$(BEDT-TTF)_2I_3$ is a very interesting problem. The difference in the skin depth and the size of each sample seems a likely explanation, but it is still under discussion and remains as a subject for future study.

In most organic conductors, smaller effective masses were obtained in the cyclotron resonance experiment than those of SdH and dHvA oscillation measurements. However, for $(BEDT-TTF)_2Br(DIA)$, the effective mass obtained by SdH measurement,[28] $4.3m_e$, is close to the mass obtained in this measurement (i.e., $\sim 4.7m_e$). Kanki and Yamada have calculated the effects of electron-electron interactions on cyclotron resonance frequency on the basis of the Fermi liquid theory.[14] They suggest that the cyclotron masses strongly depend on the characteristics of the material (such as band-filling or symmetry of FS) and in some extreme situations such as near half-filling on square lattice (four-fold symmetry), cyclotron effective masses can be comparable with the effective mass obtained by SdH or dHvA measurements. Similar to other $(BEDT-TTF)_2X$ compounds, the band-filling for $(BEDT-TTF)_2Br(DIA)$ is 3/4. Thus, this may not explain the comparable effective mass. However, recent POR results also revealed a similar effective mass value with quantum oscillation measurements,[24,29] which may suggest that the effective mass is enhanced for POR arising from the elliptical FS (two-fold symmetry) and is different from conventional CR masses.

In summary, we have performed magnetooptical measurements of $(BEDT-TTF)_2Br(DIA)$, and successfully observed POR up to the 7th order which may arise from an elliptic q2D FS. The obtained effective mass is consistent with the SdH measurements. Further angle dependence measurements are required to determine the ellipticity of the FS.

Acknowledgments
This work was supported by Grant-in-Aid for Scientific Research (B) (No.10440109), and Grant-in-Aid for Scientific Research on Priority Areas (A) (No.11136231, 12023232 Metal-assembled Complexes ) from the Ministry of Education, Science, Sports and Culture. This work has been performed at High Field Laboratory for Superconducting Materials, Institute for Materials Research, Tohoku University.

References
1) J. Singleton, F.L. Pratt, M. Doporto, W. Hayes, T. J. B. M. Janssen, J. A. A. J. Perenboom, M. Kurmoo and P. Day: Phys. Rev. Lett **68** (1992) 2500.
2) S. Hill, J. Singleton, F.L. Pratt, M. Doporto, W. Hayes, T.J.B.M. Janssen, J.A.A.J. Perenboom, M. Kurmoo and P. Day: Synth. Met. **55-57** (1993) 2566.
3) S.V. Demishev, A.V. Semeno, N.E. Sluchanko, N.A. Samarin, I.B. Voskoboinikov, V.V. Glushkov, A.E. Kovalev and N.D. Kushch: JETP Lett. **62** (1995) 228.
4) H. Ohta, Y. Yamamoto, K. Akioka, M. Motokawa and K. Kanoda: Synth. Met. **86** (1997) 1913.
5) H. Ohta, Y. Yamamoto, K. Akioka, M. Motokawa, T. Sasaki and T. Fukase: Synth. Met. **86** (1997) 2011.
6) K. Akioka, H. Ohta, Y. Yamamoto, M. Motokawa and K. Kanoda: Synth. Met. **86** (1997) 2051.
7) K. Akioka, H. Ohta, S. Kimura, S. Okubo, K. Kanoda and M. Motokawa: Physica B **246-247** (1998) 311.
8) Y. Oshima, N. Nakagawa, K. Akioka, H. Ohta, S. Okubo, M. Tamura, Y. Nishio and K. Kajita: Synth. Met. **103** (1999)1919.
9) Y. Oshima, H. Ohta, S. Okubo, K. Koyama, M. Motokawa, M. Tamura, Y. Nishio and K. Kajita: Physica B **294-295** (2001) 431-434.
10) T. Sasaki, H. Sato and N. Toyota: Synth. Met. **42** (1991) 2211.
11) P. Christ, W. Bilberacher, M.V. Kartsovnik, E. Steep, E. Balthes, H. Weiss and H. Müller: JETP Lett. **71** (2000) 303.
12) K. Miyagawa, A. Kawamoto and K. Kanoda: Phys. Rev. B **56** (1997) R8487.
13) W. Kohn: Phys. Rev. **123** (1961) 1242.
14) K. Kanki and K. Yamada: J. Phys. Soc. Jpn. **66** (1997) 1103.
15) S.J. Blundell, A. Ardavan and J. Singleton: Phys. Rev. B **55** (1997) R6129.
16) S. Hill: Phys.Rev. B **55** (1997) 4931.
17) H.M. Yamamoto, J. Yamaura and R. Kato: J. Am. Chem. Soc. **120** (1998) 5905-5913.
18) S. Uji, C. Terakura, T. Terashima, H. Aoki, H.M. Yamamoto, J. Yamaura and R. Kato: Synth. Met. **103** (1999) 1978.
19) K. Koyama, M. Yoshida, H. Nojiri, T. Sakon, D. Li, T. Suzuki and M. Motokawa: J. Phys. Soc. Jpn. **69** (2000) 1521.
20) O. Klein, S. Donovan, M. Dressel and G. Grüner: Int. J. Infrared & MMW. **14** (1993) 2423.
21) S. Donovan, O. Klein, M. Dressel and G. Grüner: Int. J. Infrared & MMW. **14** (1993) 2459.
22) M. Dressel, O. Klein, S. Donovan and G. Grüner: Int. J. Infrared & MMW. **14** (1993) 2489.
23) M. Mola, S. Hill, P. Goy and M. Gross: Rev. Sci. Inst. **71** (2000) 186.
24) J.M. Schrama, J. Singleton, R.S. Edwards, A. Ardavan, E. Rzepniewski, R. Harris, P. Goy, M. Gross, J. Schlueter, M. Kurmoo and P. Day: J. Phys. Condens. Matter **13** (2001) 2235.
25) S. Hill: Phys. Rev. B **62** (2000) 8699.
26) M.Ya. Azbel' and E.A. Kaner: Zh. Eksp. Teor. Fiz. **32** (1956) 896. [Sov. Phys. JETP **5** (1957) 730.]
27) L.P. Gor'kov and A.G. Lebed': Phys. Rev. Lett. **71** (1993) 3874.
28) S. Uji: in preparation.
29) C. Palassis, M. Mola, J. Tritz, S. Hill, S. Uji, K. Kawano, M. Tamura, T. Naito, H. Kobayashi: Synth. Met. **120** (2001) 999.

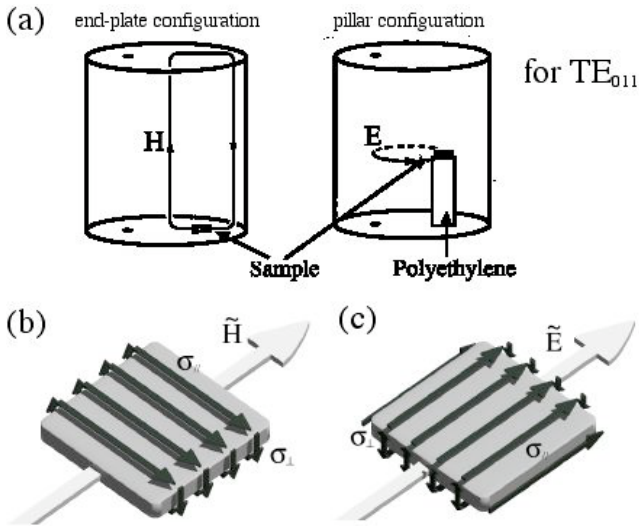

Fig. 1
(a) Sample configuration and coupling mode in the cavity (in the case of $TE_{011}$).
(b) Induced current in the case of H-field coupling.
(c) Induced current in the case of E-field coupling.

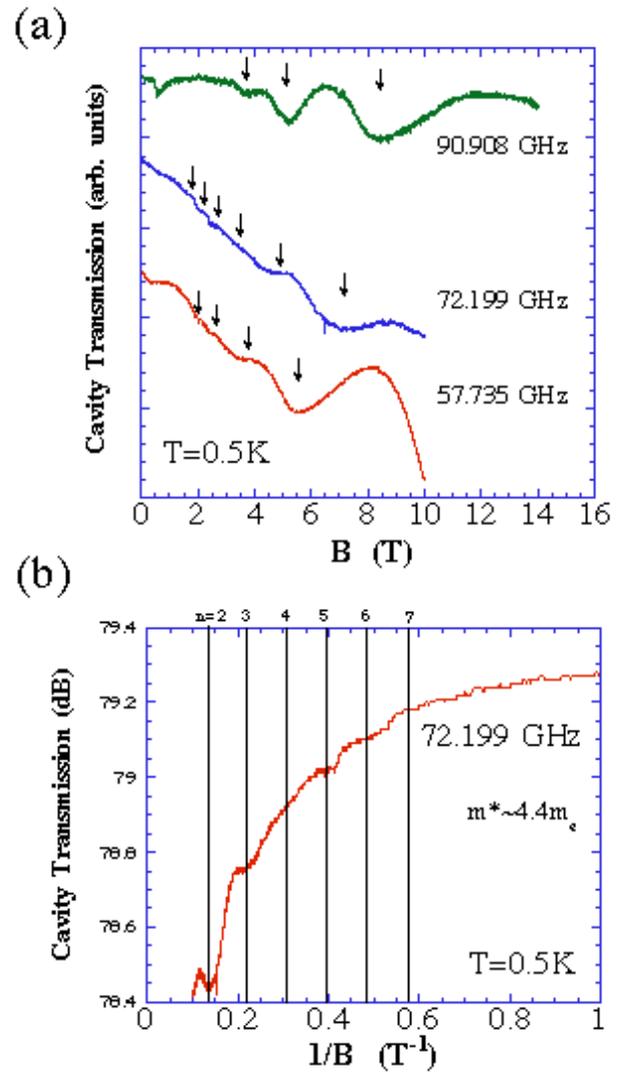

Fig. 2
(a) Typical cavity transmission spectra of $(BEDT\text{-}TTF)_2Br(DIA)$ observed at 0.5 K in the pillar configuration.
(b) Inverse-field plot of the 72 GHz spectrum. The line shows the harmonic number.

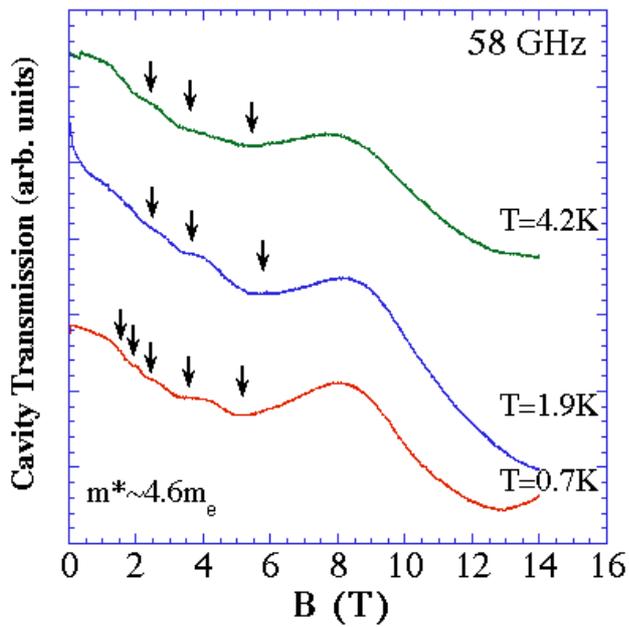

Fig. 3
Temperature dependence spectra of (BEDT-TTF)$_2$Br(DIA) at around 58 GHz.

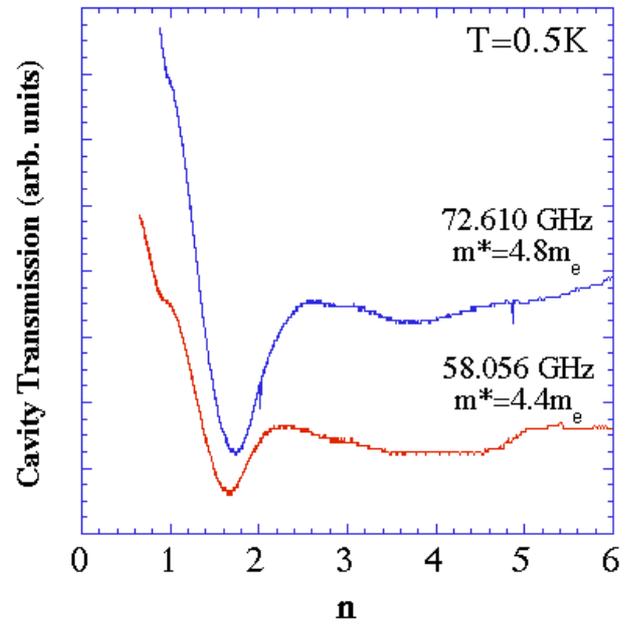

Fig. 5
Inverse-field plot of the spectrum shown in Fig. 4 (The horizontal axis is normalized by 'n' using the estimated effective mass).

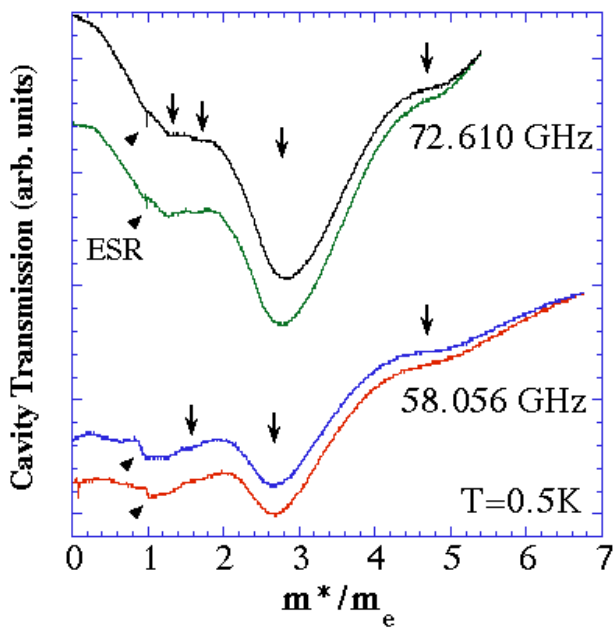

Fig. 4
Typical cavity transmission spectra of (BEDT-TTF)$_2$Br(DIA) in the end-plate configuration at 0.5 K (see text for details). The arrows show harmonics and triangles show ESR.

7